\documentclass[sigconf,authorversion]{acmart}
\acmConference[SE 2030]{International Workshop on Software Engineering in 2030}{July 2024}{Porto de Galinhas (Brazil)}

\usepackage{xspace}
\usepackage{tabularx}
\usepackage{graphicx,color}
\graphicspath{{./figs/}}
\usepackage{multirow}
\usepackage{enumitem}

\newcommand{\etal}{{\it et al.}\xspace}

\setcopyright{none}
\renewcommand\footnotetextcopyrightpermission[1]{}

\title{Some things never change: how far generative AI can really change software engineering practice}

\author{Aline de Campos}
\email{aline.campos@pucrs.br}
\orcid{0000-0002-3585-954X}
\affiliation{%
  \small 
  \institution{Pontifical Catholic University \\of Rio Grande do Sul (PUCRS)}
  \city{Porto Alegre}
  \country{Brazil}
}

\author{Jorge Melegati}
\email{jorge.melegati@unibz.it}
\orcid{0000-0003-1303-4173}
\affiliation{%
  \small 
  \institution{Free University of Bozen-Bolzano}
  \streetaddress{Piazza Domenicani 3}
  \city{Bolzano}
  \country{Italy}
}

\author{Nicolas Nascimento}
\email{nicolas.nascimento@pucrs.br}
\orcid{}
\affiliation{%
  \small 
  \institution{Pontifical Catholic University \\of Rio Grande do Sul (PUCRS)}
  \city{Porto Alegre}
  \country{Brazil}
}

\author{Rafael Chanin}
\email{rafael.chanin@pucrs.br}
\orcid{}
\affiliation{%
  \small 
  \institution{Pontifical Catholic University \\of Rio Grande do Sul (PUCRS)}
  \city{Porto Alegre}
  \country{Brazil}
}

\author{Afonso Sales}
\email{afonso.sales@pucrs.br}
\orcid{}
\affiliation{%
  \small 
  \institution{Pontifical Catholic University \\of Rio Grande do Sul (PUCRS)}
  \city{Porto Alegre}
  \country{Brazil}
}

\author{Igor Wiese}
\email{igor@utfpr.edu.br}
\orcid{}
\affiliation{%
  \small 
  \institution{Federal University of Technology \\of Paraná (UTFPR)}
  \city{Campo Mourão}
  \country{Brazil}
}

\begin{abstract}
Generative Artificial Intelligence (GenAI) has become an emerging technology with the availability of several tools that could impact Software Engineering (SE) activities. As any other disruptive technology, GenAI led to the speculation that its full potential can deeply change SE. However, an overfocus on improving activities for which GenAI is more suitable could negligent other relevant areas of the process. In this paper, we aim to explore which SE activities are not expected to be profoundly changed by GenAI. To achieve this goal, we performed a survey with SE practitioners to identify their expectations regarding GenAI in SE, including impacts, challenges, ethical issues, and aspects they do not expect to change. We compared our results with previous roadmaps proposed in SE literature. Our results show that although practitioners expect an increase in productivity, coding, and process quality, they envision that some aspects will not change, such as the need for human expertise, creativity, and project management. Our results point to SE areas for which GenAI is probably not so useful, and future research could tackle them to improve SE practice.
\end{abstract}



\date{April 2024}

\begin{document}

\maketitle

\section{Introduction}
\label{sec:introduction}

Software engineering (SE) is one of the many fields that will be affected by Generative Artificial Intelligence (GenAI). Tools such as OpenAI's ChatGPT\footnote{https://chat.openai.com}, Google's Gemini\footnote{https://gemini.google.com/app} and Github's Copilot\footnote{https://github.com/features/copilot} have become present in software developers' daily routine. This presence tends to have a profound impact on the landscape of the software development industry and brings to light the need for the research community to provide pathways forward~\cite{Ebert2023,nguyen2023generative}. 
As the general case for disruptive technologies, the expectations are not entirely fulfilled, and the real outreach of novel technology, although relevant, does not come to the extent foreseen at the emergence of the technology. For example, a Time magazine article from 1966\footnote{https://content.time.com/time/subscriber/article/0,33009,835128-5,00.html} predicted that, by 2000, ``machines will be producing so much that everyone in the U.S. will, in effect, be independently wealth.'' Even though, after almost 60 years, the automation level of our society has profoundly increased, the prophecy is far from the truth. This phenomenon also happens in SE. In the 1990s, there were expectations that CASE tools would revolutionize the industry~\cite{Vessey1992}. After more than 30 years, these expectations have not come to fruition yet and are not relevant in the industry. This issue echoes Brooks's~\cite{Brooks1987} argument that there is ``no single development, in either technology or management technique, which by itself promises even one order-of-magnitude improvement within a decade in productivity, in reliability, in simplicity.'' 

Thus, in this paper, we aim to explore this issue regarding the use of GenAI for SE by focusing on the research question: \textbf{Which aspects of SE are not expected to be deeply transformed by generative AI in the short and medium-term?} To achieve our goal, we conduct a survey with professional software developers, comparing their expectations with SE roadmaps previously proposed in SE literature. 

Our results indicate that practitioners do not expect that SE will change with respect to some aspects of requirements engineering, the need for human expertise and creativity, quality assurance, project management, and some facets of implementation and maintenance. However, they also expect positive impacts: an increased productivity, a support on analysis and development, an improved quality of the code and process, and augmented human capabilities. They also pointed out challenges to adoption of GenAI tools for SE: human and cultural issues, technical challenges, and issues regarding management and strategy, besides ethical considerations.

\section{Background and Related Work}
\label{sec:background}

Roadmaps guide researchers to focus their efforts better when conducting scientific studies. Further, they should attempt to present a vision to the research community about topics that have high potential promise to change the technology landscape \cite{kostoff2001science}. Moreover, it should be informed on what the research community is currently conscious about and effectively working to deepen understanding (the ``known knows''), what the community is aware of but not able to understand (the ``known unknowns'') and what the community is not even aware of and currently not expending effort to understand (the ``unknown unknows'') while remaining open for ``imagination of the brightest drivers of change in that field''~\cite{galvin1998science}.

Specifically for SE, roadmaps have been emerging for a long time, discussing aspects of the field and providing a path forward for researchers. For instance, in 1990, Mary Shaw~\cite{shaw1990prospects} compared SE with other consolidated engineering disciplines, observing a lack of scientific knowledge supporting the practice. Based on that and on an analysis of the field's evolution, she proposed a roadmap focusing on increasing the coupling between science and commercial practice and improving specialization and professionalization. However, collaboration between academia and industry in SE is still an issue today~\cite{Garousi2019}. \looseness=-1

In 2000, Finkelstein and Krammer~\cite{finkelstein2000software} presented a roadmap of the whole SE field, organized along activities, such as process and requirements engineering. In their work, the authors presented how SE should prepare for increased componentization, change handling, non-functional properties, developing software-as-a-service systems, critical and real-time systems, software architecture, configurability, and domain specificity. Ossher \etal\cite{ossher2000software} also presented a roadmap in the same context. In their version, focusing on tooling and environment, the main challenges rested on handling permanently malleable software, i.e., changeable, adaptable, and which incorporated concern separation. In the same year, Nuseibeh and Easterbrook~\cite{Nuseibeh2000} also proposed a roadmap on requirements engineering, pointing to further developments in modeling for analyzing the environment and capturing non-functional requirements and more formal approaches.

In 2009, Shaw \cite{shaw2009continuing} presented an outlook for which SE problems were reassembling more complex contexts, such as social contexts. Thus, the main challenge was handling the absence of system boundaries. This absence mainly referred to systems permanently connected to the internet and constantly available.

Some roadmaps in SE have made predictions or proposals for the future that did not materialize. For example, Bompani \etal \cite{bompani2000software} have proposed a roadmap in which the Extensible Markup Language (XML) format was to be widely adopted with the advent of the internet. Although XML still exists and is sometimes adopted, JavaScript Object Notation (JSON) has become more prevalent in web applications. 

Specifically towards GenAI, various works provide ideas on where the research could move towards. Ebert and Louridas~\cite{Ebert2023} have presented a roadmap that presents valuable insights from the authors about applying GenAI to increase productivity in software development, for example, creativity enhancement, summarization of documents, and problem-solving. Lo~\cite{lo2023trustworthy} has presented some key challenges concerning AI and SE, including a vision for what he called Software Engineering 2.0. In his work, the need for trust, synergy, interlink between synergy and trust, and synergy beyond trust are discussed as the main challenges. The author provides a roadmap proposing the characterization of trust factors, prevention of misplacing trust and calibrating trust, designing trust-aware efficacy metrics, building smarter systems solutions with LLM and synthesizing task user-aware explanations.

As presented above, an emphasis on what will change is a common bias of the established roadmaps. This effort could be denominated as highly speculative and open to many possibilities. Predicting how processes and tools will change over time is often difficult, especially in social contexts such as SE teams. Our study attempts to achieve a different objective, focusing primarily on what will not change from the perspective of SE. This focus should provide a basis for researchers about which of the current SE processes should continue to be investigated and further understood as they tend to continue to be required in the future of SE.

\section{Methodology}
\label{sec:methodology}

To answer our research question, we conducted a survey to gather insights, perceptions, and opinions from professional software developers about the interaction between SE practices and GenAI. To enrich this study, we strategically developed an instrument (questionnaire) to obtain different but convergent data to deepen our understanding of the research topic. Then, we analyzed the quantitative data using descriptive statistics and the qualitative data using thematic analysis. 

\subsection{Data collection}

\subsubsection{Sampling}
Initially, we selected a sample of 30 well-established IT companies known for their extensive industry reach. Selected companies either had global operations or had well-established operations in Brazil. These companies all have at least 200 employees, with the majority exceeding 10,000 collaborators and boasting an average existence of 20 years. We contacted professionals from these companies because we focused on recruiting participants with significant experience in SE. Our target audience included individuals in roles such as Chief Technology Officers (CTOs), Technical Leaders, Software Architects, Developers, and System Analysts. By targeting these professionals, we ensured data collection from individuals directly encountering the potential ramifications of GenAI integration within their work. Most of the research participants (53\%) are between 36 and 45 years old. Regarding experience working in software engineering, 47\% of the participants have between 11 and 20 years of experience. Notably, 30\% of the participants have over 20 years of experience. This distribution shows a good level of maturity and work experience. Decades of experience indicate that some participants have already experienced other periods of potentially significant changes in the field and their consequences. This scenario is representative of the software engineer population and provides a range of perspectives about GenAI practices and impressions.\looseness=-1

\begin{table}[!h]
    \centering
    \caption{Participants profile}
    \begin{tabular}{l c c}
        \multicolumn{3}{c}{Age Group} \\
         \hline
         18-25 years & &  7\% \\     
         26-35 years & & 15\% \\
         36-45 years & & 24\% \\
         46-55 years & &  2\% \\
         Over 56 years & & 0\% \\
         \hline
    \end{tabular}
    \qquad
    \begin{tabular}{l c c}
        \multicolumn{3}{c}{Work Experience} \\
         \hline
         Under 2 years & &  0\% \\ 
         2-5 years     & &  9\% \\     
         6-10 years    & & 14\% \\
         11-20 years   & & 45\% \\
         Over 20 years & & 32\% \\
         \hline
    \end{tabular}
    \vspace{-0.5em}
\end{table}

\subsubsection{Instrument}

We developed a structured online questionnaire hosted on the AirTable\footnote{https://airtable.com} platform to gather quantitative and qualitative data. The questionnaire ensured anonymity for participants and has three sections. The first section collected demographic information, including job titles, years of work experience, and age ranges. The second section assesses quantitative data collection of participants' familiarity with GenAI, its application within SE activities, and the specific GenAI tools they have tried or are using. The third section explored the perceived correlations between GenAI and SE. We referenced a series of processes in which respondents could identify which had the highest potential for GenAI integration based on their experience. We created four open-ended questions to address our objective of gathering qualitative data. We asked about their perceptions of the impacts, challenges, and ethical considerations of utilizing GenAI within SE. However, our central question was focused on aspects of SE processes that are unlikely to change significantly due to GenAI adoption.\looseness=-1

\subsubsection{Ethical Considerations}
We included an informed consent statement in the questionnaire to guarantee the participants' reliability. This statement outlined the research objectives, intended data utilization, researcher details and contact information, and data privacy assurances. The statement emphasized that collected data would be used solely for aggregated analysis and would not be disclosed individually within research outputs. Furthermore, adherence to the General Data Protection Regulation (GDPR) for data security and privacy was explicitly stated. Also, according to article 26 of CNS Resolution No. 674 of May 6, 2022 presented by the Brazilian National Research Ethics Commission, research of this nature does not require ethical assessment by the CONEP System.

\subsubsection{Pilot}

We conducted a pilot survey with five individuals to gather feedback on the instrument's design and content. This process was necessary to identify areas for improvement and ensure the quality of the data. The objective was to assess the following aspects: (a) comprehension: are the questions clear and easy to understand?; (b) bias: are there questions that could lead to biased responses?; (c) objectivity: are the questions phrased objectively? (d) cognitive load: is the questionnaire too long?; (e) language: does the language used have a pattern? The participants were asked to complete the questionnaire and provide feedback on their experience by sending text messages to the researchers, allowing immediate and direct communication. Therefore, the pilot application produced the following key findings: (a) the participants suggested that the questionnaire adopt a consistent approach, focusing on the participant's perspective; (b) there was an open-ended question regarding the benefits, however, participants indicated that we should approach GenAI's impacts to allow a more comprehensive understanding of both the positive and negative perspectives; (c) the participants expressed concerns about the potential of their responses to be linked to their employers. Based on these findings, we revised the questionnaire. We removed the mandatory question about the participant's company, considering the sensitivity of GenAI in the workplace. As suggested, we also rephrased some questions to enhance clarity and conciseness and changed the question about benefits to impacts.

\subsubsection{Execution}

Following the validation and refinement of the questionnaire, we initiated the data collection process by disseminating the survey to our target population. We opted to contact several potential respondents who matched our intended audience directly via online channels. Therefore, we avoided broad and unrestricted dissemination on social media, messaging apps, and email lists to ensure a more representative sample. We sent a brief introductory text with instructions about the questionnaire and the purpose of the data collection, emphasizing our commitment to data security and privacy. The text included a link to the online questionnaire, which was designed to be self-administered and asynchronous. The survey was available for nine days.

\subsection{Data Analysis}

We stored the 45 responses to the questionnaire in an authenticated AirTable account that only the researchers had permission to access. Initially, we did an exploratory analysis, checking for missing or incomplete responses. We identified that one participant provided only a single dot for the qualitative questions. This response was removed from the dataset, as it did not adhere to our objective to collect quantitative and qualitative data. The answers to quantitative questions were downloaded in CSV format and analyzed using Google Sheets. For single-response questions, we generated percentage distribution graphs. For multiple-choice questions, we conducted a comprehensive response count analysis. 

The answers to qualitative questions were analyzed using thematic analysis, ``a method for identifying, analyzing, and reporting patterns (themes) within data''~\cite{Cruzes2011}. It consists of reading the text, identifying specific segments of text, labeling them, translating codes into themes, and creating categories by grouping them. For coding, we employed an inductive or grounded theory approach~\cite{Cruzes2011}. This process, generally referred to open coding, can be defined as a classification process that involves a detailed examination of the data, which helps to identify initial patterns and concepts~\cite{gray2022}. It involves constant comparisons, i.e., each case identified must be compared with previous instances to determine similarities and differences. If the new case does not fit the original meaning, the definition must be modified, or a new code must be created. Therefore, this codification emerges from the data, and the analysis should occur without predetermining the relevance or expectations of previously assumed themes. We used descriptive codes to capture the central ideas or concepts in the data \cite{gray2022}. 

During the analysis process, we classified the responses into open codes, aiming to express the ideas contained within the responses. Therefore, we applied as many codes as necessary to capture different aspects within a single response. Subsequently, we grouped these into categories, intending to present themes and facilitate establishing analytical lines. One researcher initially conducted this process, and then two others reviewed the proposed coding and themes. In case of disagreement, the researchers discussed reaching a consensus. To assist in interpreting and analyzing the results, we have created some visualizations, such as graphics and tables summarizing the key information. \looseness=-1

\subsection{Supplemental package} 
The complete list of codes, their number of mentions and respective categories are available on a supplemental package on Zenodo\footnote{https://zenodo.org/doi/10.5281/zenodo.10935158}. The data was not made available in full due to privacy limitations established as a counterpart for the participation of our respondents. In this sense, the data made available begins after the researchers introduce the initial qualitative analysis. 

\section{Results}
\label{sec:results}

In this section, we present the results of our analysis. Initially, we introduce the participant's familiarity with GenAI and their perspectives on SE. Then, we examine the impacts, challenges, and ethical considerations pointed out in the qualitative data collected, presenting the main categories that emerged from the open codification of data. Finally, we discuss the categories of elements that will not change by GenAI in the participant's perspectives, as well as the insights perceived in this process.

\subsection{Familiarity with GenAI}

We inquired about the participants' familiarity with GenAI. Four options were presented for the participants to choose the one that best fit their perspective: (a) I have never heard of generative AI; (b) I know some generative AI tools; (c) I know how generative AI works, and (d) I can develop generative AI. No participants selected option (a), indicating that our sample successfully targeted professionals who are relevant in the market and usually up-to-date on the latest developments in the field. The most selected option was (c), with 44\% indicating that they are familiar with some generative AI tools. Option (d), which suggests that participants understand how this concept works, also had a good representation (37\%).

Another question asked was about the use of GenAI in their SE work. In this regard, 44\% indicated that they regularly use it and 35\% indicated they use it occasionally, not as part of their daily routine. Only 7\% indicated they are involved in developing GenAI tools, and 14\% consider themselves experts and have a decision-making role in this area. Interestingly, there is a separation between people actively engaged in GenAI, representing 56\% of the sample, and the remaining 44\% of participants are not yet actively using it.\looseness=-1 

Regarding the most used GenAI platforms, the survey found that ChatGPT from OpenAI, Copilot from Microsoft, and Gemini from Google were the most popular. The image generators DALL-E and MidJourney were also significantly mentioned. Some people also indicate using SourceAI, Tabnine, StableDiffusion, and Perplexity. This question allowed participants to select multiple responses from a list of options. We added an open field where participants could enter other examples to expand the possibilities. Other resources mentioned were AWS Codewhisperer, AWS Q, AutoGPT, Bedrock, Devin AI, Grok, Langchain, and Ollama.

These resources can offer a variety of features that make it appealing to software engineers, including the ability to generate code, translate languages, and write different kinds of content. DALL-E and MidJourney have been increasing the popularity of AI-generated art and its potential to be used in various multimedia scenarios. SourceAI and Tabnine focus on coding, indicating that some professionals are evaluating these tools' features and exploring their capabilities.

\subsection{Perspectives about SE and GenAI}

In this survey, we asked participants to identify the most commonly adopted software engineering processes and the ones where they perceive the most significant potential for GenAI.

The subject with the highest number of mentions was Coding, with 39 out of 43 participants indicating the potential of GenAI for it. Unit Testing also received a high number of mentions, with 35 participants. Around half of the participants mentioned software architecture, system testing, acceptance testing, and software maintenance. Other subjects that received several mentions were requirements elicitation, requirements validation, and requirements analysis. Change management, configuration management, systems integration, and requirements management received fewer mentions.
We also included an open-ended field for participants to suggest other relevant processes. The most frequent suggestions were software documentation, software quality, solution prototyping, and information security.

Notably, the most mentioned topics are highly repetitive and involve a lot of manual work, making them more likely to use GenAI's capabilities. Tasks such as system testing and generating software documentation can also be included here. However, more complex and technical processes, such as system architecture and complex maintenance tasks, could indicate a space for GenAI as a human partner. Analyzing the processes with fewer mentions makes it possible to verify that they involve more abstract thinking and creativity. We can infer that the participants still consider humans responsible for this task.

\subsection{Impacts}

The rise of GenAI is strongly felt across various fields of knowledge and is gradually widening its scope. Participants openly expressed their views and addressed a set of issues when asked about the main impacts of GenAI on SE. We present the categories defined in our thematic analysis for this question. \looseness=-1

\subsubsection{Agility and Productive Enhancement} This category highly references increased productivity, reduced manual/repetitive tasks, and enhanced process agility. A positive perception emerges regarding the utilization of GenAI as a tool to delegate processes currently performed by humans, yet not requiring higher-order cognitive skills. In this context, AI is perceived as a workforce capable of supporting project productivity and agility. Accordingly, some participants mentioned time optimization due to the broader adoption of GenAI. One mentioned that \textit{"there is potential yet to be explored, but it can strongly speed up the entire life cycle of software"}. Regarding potential time optimization approaches, respondents mentioned aspects of coding speed and activity automation. The potential for rework reduction was highlighted less frequently, yet still relevant. This impression shows the necessity of freeing humans to focus on more strategic tasks that require critical thinking, creativity, and complex problem-solving skills.

\subsubsection{Analysis and Development Support} Numerous routine activities in systems analysis and development were cited as strongly impacted by GenAI. The most significant number of mentions was in writing tests, indicating that GenAI can help generate unit and integration tests and even identify other scenarios for improved test coverage. Activities such as generating documentation and basic code were also pointed out. GenAI can assist in creating API documentation, insert code comments, and create user manuals easily. An interesting activity indicated was assistance with proof of concept (POC). GenAI can rapidly create prototypes based on initial requirements and automate basic functionality for POC demonstration. A participant expressed that his team \textit{"have carried out some proofs of concept and usage validations and it has been very beneficial"}. Other tasks briefly mentioned included software maintenance, migrations acceleration, data analysis support, requirements writing, and basic design activities.

\subsubsection{Process Quality Improvement} Issues were raised such as the indication of improvements in the various stages of software construction processes, especially in code review, specifications review, and fault identification. GenAI can analyze potential coding errors, style inconsistencies, and security vulnerabilities. It can also suggest best practices and provide code refactoring recommendations. A participant mentioned that they already have been using it for \textit{"code reviews and suggestions for improving commits"}. Upon analyzing the various GenAI tools that emerge daily, this has been an approach. Numerous Integrated Development Environments (IDEs) already incorporate GenAI features, such as JetBrains IntelliJ and Microsoft Visual Studio Code using Copilot. Regarding specifications, it can offer suggestions for more clarity, completeness, and consistency, potentially identifying ambiguities.

\subsubsection{Human-AI Interaction} Considering the tendency towards an increasingly collaborative process between humans and AI, general actions that assist humans were mentioned, such as aiding in human learning, providing interactive tutorials and code examples tailored to their specific needs and skill levels. In this context, a participant mentioned that \textit{"as the barrier to writing code will be lower, programmers will specialize more in other areas of computer science than in specific languages/platforms."}. Furthermore, solving simple problems and supporting the understanding of AI capabilities and limitations, will promote knowledge transfer between humans and AI. It was also mentioned that this leads to the emergence of new roles for humans and AI. Some roles can include specialists focusing on training and fine-tuning models for specific domains, AI-human interaction designers \cite{xu2023transitioning}, prompt engineers \cite{white2023prompt}, AI content reviewers \cite{leung2023best}, and AI security specialists \cite{gupta2023chatgpt}.

Although we asked about impacts, not influencing negative or positive positions, we could verify that most responses implied more positive and beneficial impacts. This perspective can lead to understanding that the practitioners are open-minded to this scenario.\looseness=-1 

\subsection{Challenges}

We asked participants their opinions about the main challenges of adopting GenAI in SE processes and divided these perspectives into three categories.

\subsubsection{Human and Cultural Issues} In this respect, the most frequently mentioned aspect was the need for human training to expand knowledge and understanding of GenAI, including model capabilities, limitations, and ethical implications. Furthermore, the necessity for human acceptance was mentioned, given that there is already a contest regarding the uses of AI and the outcomes derived from it. It is a significant challenge since the fear of job displacement and lack of trust in AI outputs have been widely discussed over the last few years. In this sense, the adaptation of developers was also highlighted, indicating the need to adjust to paradigm shifts from the increased integration of AI in SE. It will be necessary to develop new skills and adapt to new workflows. Another mention was corporate culture. Organizations that are open to innovation could be more receptive to GenAI. An interesting element cited was the potential tendency towards complacency. Other technological resources have impacted history in different ways. GenAI can lead to conformity if humans directly accept what is created without critical reasoning. Therefore, it is necessary to emphasize the importance of lifelong learning and human development enhanced by these resources and not overlapped by them.

\subsubsection{Technical Challenges} The most cited issue was information security. Concerns were expressed about the potential for sensitive data breaches, the generation of code containing confidential information, and the risk of manipulating AI models to generate malicious processes. Furthermore, the accuracy of results and the need to understand the construction of correct prompts were also mentioned. Fewer mentions were about interoperability and the potential for over engineering in cases where developers could spend excessive time and effort fine-tuning GenAI models for specific tasks. Some pointed out concerns about AI's performance and reliability in scenarios of low transparency of how the generative model works. Integration among resources was presented as an element that needs further exploration, given that a framework of technologies, platforms, and tools already exists, and there is a need to incorporate AI in convergence with these elements. As AI becomes more adopted in corporate scenarios, the increasing possibility of personalization will be necessary. Working with the specificities of scenarios and complex problems may bring significant issues regarding constructing new models and using large language models (LLMs).

\subsubsection{Management and Strategy} Some challenges addressing management and strategic contexts to incorporate AI were pointed out. The most mentioned issue is aligning the need for AI adoption with internal corporate contexts related to organizational structure, financial costs, and information security. However, there is also the difficulty of a lack of strategy in repositioning, causing a company's processes not to keep up with market movements towards the adoption of AI. Other issues also relate to the code's accountability process, as we would begin to have code produced by AI in conjunction with code produced by humans. 

Although various technical issues have been highlighted, the most critical challenges perceived by the participants are centered on human aspects, especially in developing skills to make good use of AI resources. This scenario opens discussions regarding required skills and the potential change in the profile of SE professionals. Also, it is interesting to note that these challenges presented by the adoption of GenAI could impact areas not directly related to technology, such as education, law, business, psychology, and sociology.\looseness=-1

\subsection{Ethical considerations}

The ethical implications of the heavy use of GenAI are essential to the discussion. So, we asked the participants what ethical aspects they consider essential to address when using GenAI in SE processes. After analyzing the responses, we classified three categories. Notably, five participants responded that they did not find any ethical implications or did not perceive any as truly relevant at this moment. This stance may indicate that some software engineering professionals are unaware of these aspects or do not consider them applicable to this scenario, indicating that the debate about these concerns is still in its early stages.

\subsubsection{Data Integrity and Security}

Concerns about information security are not a new topic, but the rise of GenAI has intensified this discussion. Using sensitive information in training models can create security vulnerabilities and expose confidential data. Furthermore, the evolving threat landscape requires robust security measures to protect against malicious actions that could manipulate models to produce harmful codes or outcomes. Privacy remains essential, and anonymization techniques must be employed to safeguard user data. Another recurring problem is bias in outcomes. AI outputs can reflect social biases because models are trained on data that reflects human society and culture. About this, a participant said, \textit{``some models will be created with the sole objective of being the most effective of all, without any ethical filter [..] this can become a very powerful tool in the wrong hands''}. This can lead to the perpetuation and amplification of biases, potentially resulting in discriminatory outcomes. Data transparency is also a concern, as the origin and structure of the data used to train models are usually not openly available. Moreover, GenAI can sometimes fabricate information that appears plausible but is factually incorrect. This kind of behavior has been called ``hallucination''~\cite{Ebert2023} and can mislead people to believe in inaccurate information.

\subsubsection{Legal and Regulatory} In this regard, the most mentioned concern was Intellectual Property. The ownership of AI-generated content is a complex topic. However, it is necessary to find ways to protect proprietary AI models and the innovations they represent. Patent laws may need to adapt to address AI-specific intellectual property concerns. Another consideration pointed out was human responsibility, indicating the need to be accountable for the decisions or any harm caused by AI-generated outputs. Liability must also exist in cases where AI models are misused or intentionally manipulated to produce harmful results. Some studies~\cite{Farina2024} about ethical guidelines establish principles for developing and using GenAI, essential to ensure responsible and ethical practices. The importance of compliance was also cited. For some participants, GenAI systems must comply with existing laws and regulations,\textit{``the appropriate action is to adhere to the norms established by civil and criminal codes, proceeding with the investigation and punishment of individuals who use these tools to gain illicit advantages''}. However, in some cases, it will also be necessary to adapt existing frameworks or develop new ones to supervise the development and use of GenAI technologies effectively. Another topic is the indication of human supervision throughout the AI development process, from data selection and model training to deployment and monitoring. Human judgment and intervention should also remain essential in critical decisions or potential harm.

\subsubsection{Social and Human Issues} The most mentioned topic was the potential social implications of the widespread use of GenAI. Job displacement was cited, as automation could lead to job displacement in various sectors. One participant thinks this context could \textit{``require the planning of public and union policies to prevent the precarization of work and for those who remain after the mass adoption of technology''}. This could reshape the nature of work, and it will be essential to consider how humans and AI can best collaborate to achieve optimal outcomes. An interesting topic pointed out by the participants was the tendency towards complacency. The potential over dependence on AI for tasks that could be done with human creativity and critical thinking could lead to a decline in these abilities, encouraging passive consumption rather than active engagement and critical information analysis.

\subsection{What will not change?}

When asked about what will not change in the SE area with the increasing adoption of GenAI, the participants pointed out many impressions. We classified it into the following five categories.

\subsubsection{Requirements Engineering}

Requirements engineering plays a critical role in the successful development and deployment. The most mentioned topic in this category was functional and non-functional requirements gathering. Capturing this information requires a serious understanding of the domain, project goals, and user needs. Also, requirements should be consistent, feasible, and verifiable, and the need for requirements disambiguation was pointed out. This activity often involves dealing with ambiguous or uncertain statements that must be clarified and translated into clear and actionable requirements. Emphasizing user-centered design is also important, ensuring that the software meets the needs and expectations of its intended users, especially when a project is in the discovery phase. 

Other topics included creating or following a value proposition definition for a project or product and prioritizing features and tasks. One of the participants corroborates this idea, expressing that it will be difficult to entirely change ``the core development of the product itself, such as the value proposition and requirements, things that require more creative input''. These require understanding the organization's business goals, target market, and competitive landscape, and contextual understanding and strategic insight are difficult processes for AI to perform. 

The participants demonstrate the belief that GenAI will not be flexible enough to handle dynamic situations or resolve inconsistencies in requirements. Technical constraints, resource limitations, and project timelines must be considered, and it may also not be able to capture subtleties of user behavior, preferences, and pain points.\looseness=-1

\subsubsection{Human Expertise and Creativity}

The need for experts with deep domain knowledge is especially true in complex or specialized fields where experience, critical thinking, and intuition are important. These procedures go beyond simply processing and generating data. Therefore, we must make decisions based on multiple criteria besides our expertise and experience, adapt to new circumstances and challenges, and apply knowledge and skills in various situations. This idea appears when a participant says: ``the way we operate today will be assisted by GenAI, especially in repetitive tasks [..], but the main decisions will continue to be performed by humans''.\looseness=-1

Another topic was human creativity and the capacity to break out conventional thinking patterns and develop original ideas. The nature of human creativity is often personal and reflects the individual's experiences. That is why it is necessary for empathy to help us understand and connect with other people's emotions.\looseness=-1

The relationship with clients and customers was mentioned since, in the SE field, many processes will still require building a relationship, understanding customer needs, and providing empathetic support. Besides, dealing with complex customer issues, negotiating solutions, and resolving conflicts often require human judgment and emotional intelligence.  Also, it was pointed out that collaborative human processes involve effective communication, teamwork, building consensus among different parties, and trust and relationships among team members.  

The subjective nature of some processes was also a topic. A participant mentioned that ``sometimes AI-generated ideas are too generalist, and even when explaining and trying to delve into a specific subject with the tool, it's not quite the same as a person writing and explaining the ideas.'' Subjective processes often involve understanding human values, beliefs, and cultural norms, making ethical decisions, and responding to human emotions. 

\subsubsection{Quality Assurance and Validation}

The participants highlighted the importance of some quality assurance and validation processes in software development. Some believe that, even with GenAI's capabilities, it will not replace human roles, especially in acceptance testing, which involves evaluating software against real-world scenarios and user behavior, which could involve subjective evaluation of usability, aesthetics, and user experience.

Another situation mentioned is that even with the high adoption of GenAI in the Quality Assurance field, well-structured prompts will be needed to improve the quality of AI-generated outputs. Specialists must still craft effective prompts that accurately capture the testing requirements and context to evaluate the generated outputs' correctness, completeness, and consistency. A participant said, ``even with many possible automations, the human still needs to assess whether the outcomes are as expected, and [..] evaluate whether the generated code meets the requirements and is functional.''\looseness=-1
In this sense, human review and expert validation will be important elements to ensure compliance with relevant standards, regulations, certifications, and specific business requirements.

\subsubsection{Project Management and Adaptation}

The most mentioned questions in these processes were task definition and scope management. According to the participants, the need for project managers who can break down complex projects into manageable tasks will be preserved. Understanding project goals, defining clear scope and technical requirements, and ensuring alignment among stakeholders are increasingly important.

It was mentioned that the broad adoption of agile management methods often brings unexpected challenges and quick adaptation. Another aspect of this context is the need to manage agile teams that require effective communication, conflict management, and consensus building among diverse perspectives. This scenario is already challenging to people, and we assume that it is improbable that GenAI will replace humans in this management role.
 
Also, assigning roles and responsibilities in projects requires considering people's skills, experience, and availability. It is challenging to assess these individual capabilities, delegate effectively, and ensure team members are accountable for their work.

\subsubsection{Implementation and Maintenance}

Participants' responses evidenced some specific aspects of software implementation and maintenance. The development of complex, distributed, or multi-entity systems presents non-standard situations. This intricacy demands a deep understanding of complex interactions, mixed data flows, dependencies, and potential failure scenarios to anticipate possible issues and create strong solutions. These tasks are more likely to continue as a human responsibility. 

Another frequently cited aspect was the definition of the software architecture. Participants highlighted that defining a satisfactory architecture requires a comprehensive understanding of system requirements, performance, and scalability. 

GenAI might suggest code blocks and create simple solutions. However, responses revealed other activities that still prioritize experience and human judgment, such as design construction and code customization for more complex systems, especially to preserve a well-structured codebase.

Configuration and change management also emerged as areas where human expertise is important. Effectively managing configuration files and dependencies across complex systems demands a deep understanding of system interactions and the ability to anticipate potential conflicts and risks during system updates. Finally, participants mentioned system integration challenges due to the necessity of aggregating different software components and knowledge of specific APIs, data formats, and communication protocols.

\begin{table}[!h]
    \caption{Number of codes per category}
    \scalebox{0.75}{
        \begin{tabular}{l c}
            \multicolumn{2}{c}{IMPACTS OF GEN AI} \\
             \hline
             Analysis and Development Support   & 9 \\     
             Agility and Productive Enhancement & 7 \\
             Human-AI Interaction               & 5 \\
             Process Quality Improvement        & 5 \\
             & \\
        \end{tabular}
    }
    \scalebox{0.8}{
        \begin{tabular}{l c}
            \multicolumn{2}{c}{ETHICAL ISSUES} \\
             \hline
             Data Integrity and Security & 5 \\     
             Legal and Regulatory        & 5 \\
             Social and Human Issues     & 4 \\
             & \\
             & \\
        \end{tabular}
    }
    \quad
    \scalebox{0.75}{
        \begin{tabular}{l c}
            \multicolumn{2}{c}{WILL NOT CHANGE} \\
             \hline
             Requirements Engineering           & 7 \\     
             Human Expertise and Creativity     & 6 \\
             Quality Assurance and Validation   & 5 \\
             Project Management and Adaptation  & 4 \\
             Implementation and Maintenance     & 10 \\
        \end{tabular}
    }
    \scalebox{0.8}{
        \begin{tabular}{l c}
            \multicolumn{2}{c}{CHALLENGES TO ADDRESS} \\
             \hline
             Technical Challenges       & 11 \\     
             Human and Cultural Issues  & 5  \\
             Management and Strategy    & 4  \\
             & \\
             & \\
        \end{tabular}
    }
    \vspace{-1em}
    
    \label{tab:table2}
\end{table}

\section{Limitations}
\label{sec:limitations}

This section presents the limitations of our study, concerning construction and internal validity, external validity, and reliability, which may influence the interpretation of the findings \cite{runeson:2009}. Construction validity: 
a common concern in this regard for qualitative studies is if the subjects had the same understanding of the questions as the researchers. To mitigate this issue, we performed a pilot and improved the questionnaire accordingly.
Internal Validity: while the systematic approach aimed to minimize biases and errors in synthesizing the results from the survey, it is possible that researchers might have been influenced to find concepts that should not change with the adoption of GenAI in SE. To mitigate this aspect, initially, the researchers conducted the analysis steps separately and independently. After the results were obtained, other researchers verified the analysis and contributed to it as well. Further, doubts were discussed synchronously and asynchronously until a consensus was reached. External validity: the generalization of our findings is constrained by the limited sample of practitioners that answered the questionnaire. However, our results point to some interesting results that could be further explored in future work which should aim for results generalizable for other contexts. Reliability: the reproducibility of this study's findings may be influenced by the dynamic nature of SE and the implications that AI could have on it, which are continually evolving. To mitigate this, we have provided all the parameters applied in our survey and ad-hoc literature review so that other researchers can replicate it.

\section{Conclusions and Future Work}
\label{sec:conclusion}

In this investigation of practitioners' expectations of GenAI for SE, we focused on the nuanced dynamics between emerging technological innovations and the foundational pillars of SE. Our survey of SE practitioners has shed light on a landscape where anticipation and realism may not converge, highlighting SE aspects that could never change. Our findings underscore a pivotal insight: while practitioners expect that GenAI will augment aspects of SE with significant enhancements in productivity, code quality, and process efficiency, there are intrinsic elements of SE that, according to the respondents, will remain impervious to automation's touch. Human creativity, domain-specific expertise, and ethical judgment emerge as indispensable pillars, highlighting the irreplaceable value of human intellect in navigating complex, nuanced, and ethically charged decisions. This duality presents a future where GenAI collaborates, acting as a coadjuvant -- not the protagonist, augmenting human capabilities rather than displacing them.\looseness=-1 

As future work, we plan to deepen our analysis, using the insights we have obtained from this study to guide the execution of a Delphi study~\cite{Okoli2004} with experts in the field. The goal of the Delphi method is to obtain the most reliable consensus of a group of experts~\cite{Okoli2004}. By crossing the input received from experts in the field of SE and GenAI with the findings of this study, we aim to identify areas that could be neglected in the following years for not being suitable for GenAI but that should be addressed. Researchers could use this list to propose novel approaches in these areas. Finally, this understanding can support the creation of new guidelines for SE education in the context of ever-growing GenAI adoption.

\balance

\begin{acks}
This study was financed in part by the Coordena\c{c}\~{a}o de Aperfei\c{c}oamento de Pessoal de N\'{i}vel Superior - Brasil (CAPES). Igor Wiese thanks CNPq \#408812/2021-4,  MCTIC/CGI/FAPESP \#2021/06662-1, Fundação Araucaria and UTFPR. This study was also partially supported by the Ministry of Science, Technology, and Innovations from Brazil, with resources from Law No. 8.248, dated October 23, 1991, within the scope of PPI-SOFTEX, coordinated by Softex.
\end{acks}

\bibliographystyle{ACM-Reference-Format}
\bibliography{Biblio}

\end{document}